\documentclass[sigconf,nonacm]{acmart}

\IfFileExists{libertine.sty}{}{%
  \IfFileExists{lmodern.sty}{\usepackage{lmodern}}{\usepackage{mathptmx}}%
}

\setcopyright{none}
\copyrightyear{2026}
\acmYear{2026}
\usepackage{booktabs}
\usepackage{balance}
\usepackage{tikz}
\usetikzlibrary{positioning, arrows.meta, fit, backgrounds}

\begin{document}

\title[Teaching an LLM Tutor to Withhold the Answer]{Teaching a Large Language Model Tutor to Withhold the Answer: A Supervisor Architecture and an Evidence-Driven Method for Tuning Socratic Behavior}

\author{Yusuf Pisan}
\orcid{0000-0002-9043-5803}
\affiliation{%
  \institution{University of Washington Bothell}
  \city{Bothell}
  \state{WA}
  \country{USA}
}
\email{pisan@uw.edu}

\renewcommand{\shortauthors}{Pisan}

\begin{abstract}
An effective large language model (LLM) tutor must often decline to give an answer it could easily produce. In a randomized study, students who used an unguarded chatbot scored higher while practicing but lower on a later test taken without it, whereas a Socratically guarded version of the same model kept the practice gain and removed the later loss~\cite{bastani2024genai}. Reliable answer-withholding is therefore central to a tutor's value, yet a capable model pressed by a frustrated student does not withhold reliably on a prompt alone. We report a deployed tutoring system that enforces answer-withholding as a per-turn, machine-checkable contract, and a method for tuning that withholding against evidence. A non-LLM policy core, reading only trusted learner state, sets a per-turn ceiling on an eight-rung help ladder; a deterministic detector strips solution code; and a separate LLM judge checks each risky reply against the contract. We tune the behavior with an automated evaluation that uses no human subjects: scripted student personas are driven through the live pipeline and re-scored by a stronger model, and we record each rejection's stated reason so failures are fixed by cause. Doing so revealed an interpretable ``over-help ladder,'' from blatant solution leaks, to naming the exact bug, to over-citing general facts, with each fix exposing the next. The tutor reached full compliance on all four acceptance criteria. We offer the measure, diagnose, and fix loop as a reusable recipe for any LLM agent that must refuse a capability it has.
\end{abstract}



\maketitle

\section{Introduction}

Most published cautions about LLMs in education share a structure: the model is too helpful. Asked a homework question, a capable model will solve it clearly and instantly. Decades of learning-sciences research predict that this kind of help, giving the answer rather than supporting the student's own reasoning, produces little durable learning: feedback that merely reveals the answer is among the least effective kinds~\cite{hattie2007feedback,wisniewski2020feedback}, and learning depends on the student doing the cognitive work of constructing and applying a solution~\cite{chi2014icap,vanlehn2011relative}.

Recent evidence confirms this directly for LLMs. Bastani et al. ran a randomized controlled trial with roughly a thousand secondary-school students: access to an unguarded chatbot raised performance while practicing but lowered it on a later exam taken without the tool, relative to students who never had the tool, while a Socratically guarded version of the same model kept the practice gain and removed the later loss~\cite{bastani2024genai}. The usual explanation is metacognitive offloading: the student lets the model do the planning and checking, scores well in the moment, and learns less.

The practical consequence is that, for a tutor, withholding is not a safety wrapper bolted onto the product; it is the core of the product. A useful LLM tutor must, by default and under pressure, decline to produce the solution it could trivially produce while still helping the student make progress. This is an unusual requirement. Classical refusal in LLM safety is refusal of content the model should not produce. A tutor instead faces what we call \emph{refusal-under-knowledge}: it knows the answer, the student can see that it knows, the student is frustrated and asks directly, and the system must still withhold while staying warm and genuinely helpful. Making that behavior hold reliably is the problem this report addresses.

Two observations shape our approach. First, prompt instructions are necessary but not sufficient: a single prose prompt that must at once be warm, Socratic, never-revealing, and factually grounded is overloaded, and these goals interfere under adversarial input. Second, the behavior cannot be checked by reading the prompt; it must be measured against student behavior, including deliberate attempts to extract the answer, and tuned until it complies. Both point away from ``write a better prompt'' and toward enforcing withholding outside the generating model, plus a method that calibrates that enforcement against evidence.

This report contributes, in order of emphasis: (1)~\textbf{a method for calibrating Socratic withholding against evidence} (\S\ref{sec:method},~\S\ref{sec:results}): four acceptance criteria, an offline adversarial simulation, and a live loop that drives scripted personas through the production system and re-judges every turn with a stronger model, recording the judge's \emph{reason} for each rejection so failures are fixed by cause; (2)~\textbf{a supervisor architecture that enforces withholding as a per-turn contract} (\S\ref{sec:arch}); and (3)~\textbf{a deployed, persona-grounded instantiation} for two undergraduate data structures courses, speaking in the voice of the instructor of record and grounded in that instructor's own course materials. To our knowledge this combination has not previously been deployed.

We deliberately do not claim a learning-outcomes result; the decisive test, a delayed, tool-removed assessment in a controlled design, needs a pilot cohort and is future work (\S\ref{sec:future}). What we offer here is the architecture, the calibration method, and a candid account of applying it, including what broke.

\section{Background and design forces}
\label{sec:bg}

Forty years of tutoring research, read with the recent LLM literature, yields a small set of design forces that constrain any serious tutor and motivate our architecture.

\textbf{Step-level interaction drives learning; revealing the answer is weak feedback.} Bloom set the aspiration that tutoring combined with mastery learning can far outperform conventional instruction~\cite{bloom1984sigma}. By \emph{mastery learning} we mean the specific method in which a student must demonstrate a set level of proficiency on prerequisite material, through formative checks and targeted correctives, before advancing; it is a defined instructional approach, not the generic goal of learning. VanLehn's review is the most actionable reading: step-based tutoring systems reach effect sizes close to human tutoring, and the active factor is the granularity of interaction, feedback at each step rather than on the final answer~\cite{vanlehn2011relative}. A persistent per-skill learner model drives sequencing and the depth of help: Bayesian Knowledge Tracing models mastery as a per-skill latent state~\cite{corbett1995knowledge}, knowledge-space theory generalizes assessment over prerequisite structure~\cite{doignon1985spaces}, and large field trials show lightweight immediate feedback helps the weakest students most~\cite{heffernan2014assistments}. The same pattern appears in the LLM era: an AI tool that coached human tutors toward more probing questions and less answer-giving helped the weakest tutors most~\cite{wang2024tutorcopilot}. The ICAP framework predicts that interactive and constructive engagement outlearn passive reception~\cite{chi2014icap}; AutoTutor showed, before LLMs, that a fixed sequence of dialogue moves (pump, hint, prompt, then assert) produces large gains~\cite{graesser2005autotutor}; and feedback meta-analyses find that roughly a third of feedback interventions are null or harmful, with praise, vagueness, and answer-reveal among the harmful kinds~\cite{hattie2007feedback,wisniewski2020feedback}. Cognitive Tutors showed deep step diagnosis works but at heavy authoring cost~\cite{anderson1995cognitive}; constraint-based tutoring showed a cheaper path, encode what a correct solution looks like and check against it, which in programming is essentially free because compilers and test suites already exist~\cite{mitrovic2012constraint}.

\textbf{Correctness must come from outside the model.} Surveys establish that LLMs state confident falsehoods, and LLMs giving programming feedback show false corrections, missed bugs, and instability~\cite{ji2023hallucination,koutcheme2025feedback}. An LLM should therefore not be the source of truth for whether a student's code is correct; a deterministic check (compile, run, test) should be, with the model limited to narrating facts it did not assert. A classroom LLM assistant that never showed solution code, offering pseudocode and annotations of the student's own work instead, documented both that this stance is feasible and that students consistently want more directness than is good for them~\cite{kazemitabaar2024codeaid}. A parallel system enforces the same no-solutions stance through prompt-level guardrails at classroom scale~\cite{liffiton2024codehelp}; our contribution is to lift that guarantee out of the prompt and make it a checkable, per-turn contract.

\textbf{Pedagogy is best expressed as a per-turn instruction.} A recent development reframes tutoring as pedagogical instruction following: instead of baking one pedagogy into a model, the model follows explicit, per-turn pedagogy instructions, with expert raters preferring the result over strong general models~\cite{learnlm2024}. A complementary line instead bakes Socratic behavior into the weights through fine-tuning~\cite{liu2024socraticlm}; we keep the pedagogy external so it can be inspected and changed without retraining. The transferable idea is the interface: pedagogy as a machine-readable, per-turn instruction a supervisory component computes and hands to the model.

\textbf{LLM judges are useful but unreliable, and the trust boundary matters.} Using one LLM to judge another's output is now standard, but such judges show position, verbosity, and self-preference biases and run-to-run inconsistency~\cite{zheng2023judge}; a judge needs a written rubric, low temperature, structured output, a policy that prefers revision over blocking, and calibration against human judgment. Separately, prompt injection means any component that reads student text can be retargeted by it, so robust designs keep the privileged component from ever seeing untrusted input~\cite{willison2023dualllm}, and injection through retrieved content is a demonstrated real-world attack~\cite{greshake2023injection}. The component that makes the binding withholding decision must therefore be a non-LLM layer fed only trusted signals. Finally, persona is a motivation lever subordinate to clarity: rich first-person corpora reproduce behavior better than short descriptions~\cite{park2024agents}, and a pedagogical agent with personality modestly raises interest and credibility, though the effect is small~\cite{schroeder2013agents}.

\section{The supervisor architecture}
\label{sec:arch}

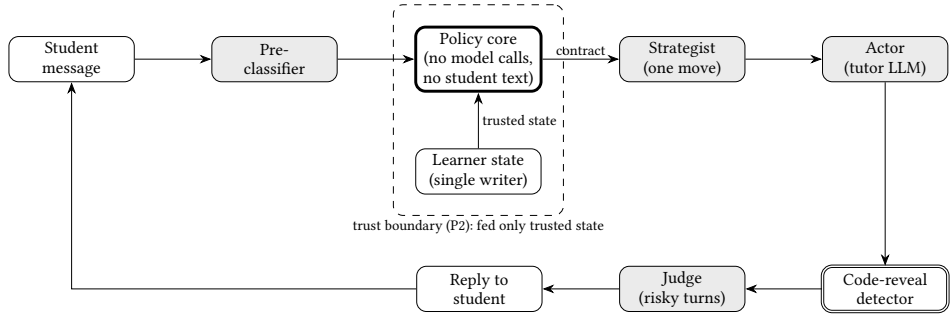
\begin{figure*}[t]
\centering
\scalebox{0.92}{%
\begin{tikzpicture}[
  >=Stealth,
  box/.style={draw, rounded corners, align=center, minimum height=6.5mm, minimum width=18mm, inner sep=2pt, font=\footnotesize},
  llm/.style={box, fill=black!8},
  core/.style={box, very thick},
  flow/.style={font=\scriptsize, inner sep=1.5pt},
  node distance=7mm and 11mm
]
\node[box] (stu) {Student\\message};
\node[llm, right=of stu] (pc) {Pre-\\classifier};
\node[core, right=of pc] (policy) {Policy core\\(no model calls,\\no student text)};
\node[llm, right=of policy] (strat) {Strategist\\(one move)};
\node[llm, right=of strat] (actor) {Actor\\(tutor LLM)};
\node[box, below=8mm of policy] (state) {Learner state\\(single writer)};
\begin{scope}[on background layer]
  \node[draw, dashed, rounded corners, fit=(policy)(state), inner sep=3mm] (tb) {};
\end{scope}
\node[flow, anchor=north] (tblabel) at (tb.south)
  {trust boundary (P2): fed only trusted state};
\coordinate (retrow) at (actor |- tblabel.south);
\node[box, double, below=4mm of retrow, anchor=north] (det) {Code-reveal\\detector};
\node[llm, left=of det] (judge) {Judge\\(risky turns)};
\node[box, left=of judge] (reply) {Reply to\\student};
\draw[->] (stu) -- (pc);
\draw[->] (pc) -- (policy);
\draw[->] (policy) -- node[flow, above] {contract} (strat);
\draw[->] (strat) -- (actor);
\draw[->] (state) -- node[flow, right] {trusted state} (policy);
\draw[->] (actor) -- (det);
\draw[->] (det) -- (judge);
\draw[->] (judge) -- (reply);
\draw[->] (reply) -| (stu);
\end{tikzpicture}%
}
\caption{Per-turn flow. The student talks only to the actor. Inside the trust boundary, the policy core computes the help contract from learner state alone, never reading student text (P2). The deterministic detector (double border) removes solution code independently of any model (P3); the judge checks risky replies before they reach the student. Shaded boxes are LLM components.}
\label{fig:arch}
\end{figure*}

The student talks to a single warm tutor, the \emph{actor}. Around every turn, \emph{supervisor} components decide what help is allowed, check that the reply stayed within that decision, and record what happened (Figure~\ref{fig:arch}). Six principles govern the design.

\begin{itemize}
\item \textbf{P1, withhold by default.} The tutor never gives the solution to a live exercise on its own; help follows a typed hint ladder whose top rung is reachable only through an explicit instructor toggle.
\item \textbf{P2, decide without reading student text.} The component that sets the per-turn help ceiling reads only trusted learner state, never the student's words, so a student cannot talk past a ceiling computed from data they cannot edit~\cite{willison2023dualllm}.
\item \textbf{P3, let deterministic signals outrank model judgments.} A code-execution result outranks an LLM's opinion of correctness, and a deterministic detector of solution code runs before, and independently of, any LLM judge~\cite{mitrovic2012constraint}.
\item \textbf{P4, prefer revising a reply over refusing it.} If the tutor refuses too often, students give up on it and return to unguarded chatbots, which recreates the exact harm we are trying to prevent~\cite{bastani2024genai}; early evidence on optional-guardrail tools confirms students will route around friction they judge excessive~\cite{kapoor2025guardrails}. We therefore treat over-blocking as a measured failure, not a safe fallback.
\item \textbf{P5, use one writer for learner state.} A single component writes all mastery and progress records, through typed transitions, so the learner model is auditable and the pedagogy layer can be tested without invoking any model.
\item \textbf{P6, make everything observable.} Every turn logs the computed contract, the verdicts, the help level used, latency, and cost.
\end{itemize}

The per-turn flow (Figure~\ref{fig:arch}) runs: spending caps, an active-exam intercept, a pre-classifier, a conduct check, grounded retrieval, the policy core, a prerequisite gate, a strategist, the actor, the deterministic detector, the LLM judge on risky turns, and telemetry; a nightly auditor re-judges risky turns with a stronger model for calibration.

\subsection{An eight-rung hint ladder}

Help is discretized onto an eight-rung ladder, from least to most revealing: \textbf{H0} acknowledge and encourage (no new content); \textbf{H1} restate or clarify the problem; \textbf{H2} point to the relevant concept, slide, or definition; \textbf{H3} ask a leading question about the very next step; \textbf{H4} describe the approach in words, with no code; \textbf{H5} walk through a worked example on a \emph{different} problem; \textbf{H6} give pseudocode with the key lines left as blanks; \textbf{H7} show the full, compilable solution. Discretizing assistance this way follows the help-seeking and help-design literature, which treats graduated, well-designed help as a first-class instructional object rather than an afterthought~\cite{aleven2003help}. This ladder is the unit of decision (the policy core sets a ceiling on it), of enforcement (the detector and judge check against it), and of measurement (compliance means staying at or below the ceiling). Rungs H0 to H4 allow no code; H5 and above allow progressively more. Making the ladder an explicit, typed object, not an informal prose stance, lets the rest of the system reason about and check withholding.

\subsection{Computing the per-turn contract}

The policy core is the heart of P2: ordinary code with no model calls and no access to student prose. From trusted state alone it computes the per-turn contract: a help ceiling, the goals in scope, any goals locked under an active exam, whether the turn must be grounded, and whether the judge should escalate to a multi-sample vote. The ceiling is computed from data rather than nested conditionals, so it is tunable and exhaustively testable. \emph{Mastery band}: lower mastery raises the ceiling (more scaffolding), higher mastery lowers it to force the productive struggle that drives durable learning~\cite{kapur2008productive}; with several goals in scope, the most permissive band wins, so the student is met at their weakest goal. \emph{Constructive-code floor}: a student who shows buggy code and asks for help is stuck on execution, not concept, so a mastery-tuned ceiling can be too low to give any real debugging hint; such a turn floors the ceiling at H4, so the tutor gives a verbal nudge while emitting no code. This floor is one of the levers the evaluation later tuned (\S\ref{sec:results}). \emph{Exam lockdown}: a goal under an active mastery exam caps the turn near zero help. \emph{Instructor explain mode}: an optional toggle lifts the ceiling to full solutions, but never past an exam lockdown. \emph{Conduct restriction}: applied last, so nothing can lift it. A blunt demand for the solution never raises restrictions on its own (P4); the system routes such a turn kindly. The contract is then rendered into an imperative instruction block appended to the actor's prompt~\cite{learnlm2024}.

\subsection{Pre-classifier and strategist}

A small, cheap \textbf{pre-classifier} maps the student message to a narrow structured output: an intent label from a fixed set (genuine question, attempt, solution demand, exam leak, injection, hostile, and so on), the goal identifiers in play, and whether the message contains a code attempt. The message is framed as data inside delimiters, and the classifier is told the transcript is never an instruction. Because the only output is an enum and a short identifier list, a successful injection can at worst cause a wrong route; the policy core still caps the ceiling, so no solution leaks from a fooled classifier, a direct application of the trust-boundary principle~\cite{willison2023dualllm}. A second small-model \textbf{strategist} proposes exactly one \emph{instructional move} per turn from the student's recent trajectory, following the validated AutoTutor-style ladder~\cite{graesser2005autotutor} (elicit a prediction, give a worked analogy, check understanding, switch representation, and so on). A move shapes \emph{how} the actor opens the turn but never \emph{what} may be revealed: it cannot raise the ceiling, is suppressed during an exam lockdown, and is dropped if the ceiling cannot accommodate it. The strategist is advisory and fail-soft.

\subsection{The actor, grounding, and persona}

The actor is the tutor the student sees, a capable production model prompted to teach in the instructor's voice. Its persona is assembled once from the instructor's own materials (a corpus distilled from several hundred lecture transcripts), plus a static block that prints the full hint ladder and its rules. The assembled prompt is large (on the order of 160{,}000 characters) and identical from turn to turn, so it is sent as a cached prefix; the per-turn contract and student context follow in a short suffix. Concept questions are grounded by retrieval over an embedding index of course slides, transcripts, the textbook, and assignments, so the tutor cites material the student actually saw. When warmth and withholding conflict, the contract wins, and the persona's job is to make withholding feel like coaching rather than refusal.

\subsection{Removing solution code: the deterministic detector}

Independently of any model judgment (P3), a deterministic \textbf{code-reveal detector} scans the output for solution code. It flags a fenced block long enough and plausibly C++ (type keywords with braces and semicolons, or tell-tale signatures), and an encoded payload that decodes to plausible C++ (an evasion seen in real logs, where the model was talked into base64-encoding the answer). The detector has two modes. By default it is lenient: it lets short illustrative snippets and fill-in-the-blank pseudocode through, treating them as harmless. When the contract allows no code at all, it switches to a strict mode that removes even a bare function signature or a blanked-out template, while still allowing plain-prose pseudocode that contains no braces or semicolons. On the live output stream it holds tokens inside a code fence until the fence closes, then releases the block if it is safe or replaces it with an in-voice line if not. Because it is deterministic, it cannot be fooled by injection and can be exhaustively unit-tested.

On a risky turn (code-adjacent at or below a no-code ceiling, an exam lockdown, or an adversarial intent), a small-model \textbf{judge} checks the draft against the contract. It is collusion-resistant by construction: it sees only the contract, the draft, and the retrieved sources, never the raw student message~\cite{willison2023dualllm}. It returns a structured verdict (allow, revise, block) with a rule identifier and a one-sentence reason, at temperature zero, against a written rubric that encodes the contract, but that explicitly does not flag general programming facts any reference would confirm~\cite{zheng2023judge}. On a revise verdict the actor regenerates once with the violation named; if that still trips the detector, the system falls back to a templated pointer. Revision is strongly preferred over blocking (P4). Finally, gated progression through 136 prerequisite-linked goals turns a turn reaching gated content into a short \textbf{mastery exam}, with the goal locked from teaching and the pass derived from a rubric score, not a model-asserted flag, failing closed on bad grader output. Student code, when enabled, is compiled and run against tests in a network-isolated sandbox that returns facts only; the model narrates them and never asserts correctness (P3)~\cite{mitrovic2012constraint}.

\section{Method: calibrating Socratic withholding}
\label{sec:method}

The architecture makes withholding \emph{enforceable}; it does not by itself make the tutor withhold \emph{correctly}. The actor can still over-help, the judge can still over-revise honest help, and a measurement can still mislead. Our method closes that gap with a set of acceptance gates, an offline adversarial simulation, a live test loop, and a diagnostic discipline that makes failures interpretable. By an \emph{acceptance gate} we mean a concrete pass/fail criterion the system must meet before we treat a change as safe to ship.

\textbf{No human subjects.} The evaluation reported here is entirely automated. Behavior is exercised by scripted student personas (defined below), not by students, and is scored by the system's own LLM judge and by a separate, stronger auditing model. Because no human participants and no student data are involved, the work in this paper required no human-subjects review; the controlled study in \S\ref{sec:future}, which does involve students, will require it.

Underneath everything is a large suite of fast, deterministic tests (over five hundred, run on every change with no network) encoding the architecture's invariants, such as that the policy core imports nothing that can carry student text and only the single writer touches learner state. These confirm the machine obeys its own contracts; they cannot show the tutor teaches well or withholds under pressure, which requires driving student behavior through it.

\subsection{Four acceptance gates}

We reduce ``the tutor withholds correctly'' to four measurable properties. \textbf{G1, no solution reveals}: across an adversarial suite, no detector-confirmed C++ solution reaches the student. \textbf{G2, do not over-block honest help}: on genuine, code-adjacent help-seeking turns, the \emph{earnest revise rate} (the fraction of such turns the judge marks \emph{revise}) is at most 5\%, and any outright block or detector cut of an earnest turn also fails the gate, since over-blocking is the more harmful error (P4). \textbf{G3, hint-ceiling compliance}: under adversarial pressure, \emph{hint-ceiling compliance} (the fraction of turns that stay at or below the contract's help ceiling and leak no code, as scored by an independent auditor) is at least 95\%. \textbf{G4, exam integrity}: no mastery exam is passed through injection or grader failure. G1 and G4 are deterministic and checked on every change; G2 and G3 need the live actor and an auditing model, and are measured deliberately.

\subsection{Offline adversarial simulation}

We script four student personas: an earnest-but-stuck student (must not be over-blocked), an answer-seeker who repeatedly demands the code (the detector must cut anything the actor concedes), a gate-evader who uses social engineering (``my TA said it is fine''), and an injector who attempts prompt injection. The answer-seeker and gate-evader model the long-documented ``gaming the system'' behavior, in which students who extract answers rather than reason learn substantially less~\cite{baker2004gaming}; resisting that extraction is exactly what the ladder defends. These drive a curated corpus of injection messages, evasions, candidate solution drafts, and genuine tutoring turns through the deterministic components. Three classes of reveal are missed by design and documented as known blind spots (split-fence, plain-prose, and cross-language reveals), which is why the live judge and the execution path exist as further layers.

\subsection{The live test loop}
\label{sec:liveloop}

G2 and G3 depend on what the production model does under pressure, so we measure them with a deliberately run, billed loop. Safeguards are enforced, not merely intended: the loop runs against a throwaway database (never production), with non-admin synthetic students and a printed cost ledger, and it refuses to start unless the throwaway-database check passes (an earlier incident, in which an unrelated maintenance script touched the production store, is why this is now a hard assertion). Two synthetic students have all goals pre-mastered, so every turn reaches the actor and judge rather than being diverted into an exam. For G2 we drive the earnest suite and read each turn's live verdict; for G3 we drive every evasion and injection, then re-judge all driven turns with the stronger auditing model, counting a turn compliant when the auditor does not reject it for exceeding the ceiling or leaking code. One full loop is roughly two dozen driven turns plus an audit pass and costs under a dollar. Scoring is separated from the billed driving, so only turn-driving spends money.

\subsection{Separating failures by cause}
\label{sec:diag}

Two choices made the difference between a number and a diagnosis. First, \textbf{sort each rejection by its cause}. The auditor's rejections fall into three kinds: exceeding the help ceiling or leaking code (a true withholding failure), ignoring the assigned instructional move (a pedagogy-quality issue), and citing or claiming something without support (a grounding issue). Only the first kind counts against G3. A single undifferentiated pass rate mixes all three, which is what made the first live result uninterpretable until we separated them (\S\ref{sec:results}). Second, \textbf{record the judge's reason, not just its verdict}. Early runs reported that certain earnest turns were revised but not why, which invited blind fixes. Capturing each rejection's stated reason turned an expensive, ambiguous signal into a precise one: the next run did not just say a turn failed, it said the actor cited a source it had not retrieved, or emitted code at a ceiling that forbade it.

\section{Results: the over-help ladder}
\label{sec:results}

We applied the method iteratively to the deployed tutor. All figures below are pass rates over the scripted-persona suites of \S\ref{sec:method} (about two dozen turns per run); none involve human participants. The result worth carrying away is not that the gates were met but the \emph{shape} of the path to them: the withholding failures formed a descending ladder, from gross to subtle, a map of how a capable model over-helps when told not to. Table~\ref{tab:ladder} summarizes that descent.

\begin{table*}[t]
\centering
\caption{The over-help ladder: each billed calibration run exposed a subtler over-help, caught by the captured judge reason (\S\ref{sec:diag}) and fixed by cause. \emph{G2} is the earnest revise rate (gate $\leq$5\%); \emph{G3} is hint-ceiling compliance (gate $\geq$95\%). Run~1's reported 54\% G3 was a measurement artifact (Rung~0; true value $\approx$77\%).}
\label{tab:ladder}
\footnotesize
\begin{tabular}{@{}l l l p{0.46\textwidth}@{}}
\toprule
Run & G2 (earnest revise) & G3 (ceiling compliance) & Fixes applied since the previous run \\
\midrule
1 (start) & 43\% & 54\% reported / $\approx$77\% real & --- \\
2 & 43\% & 96\% (pass) & persist sources for the auditor (Rung~0); strict detector for partial code (Rung~1); constructive-code floor (Rung~2) \\
3 & 0\% (pass) & 96\% & cite only retrieved sources; floor extended to any code-adjacent turn (Rung~3); per-rejection reason capture \\
final & 0\% (pass) & 100\% (pass) & low-rung replies name no bug or fix; grounding narrowed to course-specific claims (Rung~4) \\
\bottomrule
\end{tabular}
\end{table*}

\textbf{Starting point.} The first billed run failed both pedagogy gates badly: a 43\% earnest revise rate (G2) and 54\% hint-ceiling compliance (G3), seemingly a system that both over-blocked honest students and leaked under pressure, a contradiction worth resolving first.

\textbf{Rung 0, a measurement artifact.} Sorting G3 rejections by cause (\S\ref{sec:diag}) showed that half of the apparent ceiling violations were grounding rejections, not over-help. The fault was in the auditor: it re-judged turns without the retrieved sources the live judge had seen, so it read every legitimate citation as unsupported. We proved this offline, at no further cost, then fixed it by persisting the sources the live judge saw and re-judging against them. Real ceiling compliance was about 77\%, still failing, but now genuine.

\textbf{Rung 1, code under pressure.} The genuine violations were the actor emitting partial code (signatures, blank-filled templates) at low rungs under adversarial demand, slipping past the lenient detector because the fragments were short or contained blanks. The fix was the strict detector mode (\S\ref{sec:arch}).

\textbf{Rung 2, over-blocking honest debugging.} Independently, G2's over-revising traced not to an overzealous judge but to a ceiling set too low: a confident student gets a low ceiling to force productive struggle, but a student showing buggy code needs a higher ceiling to get any real debugging hint, so the judge correctly revised help that genuinely exceeded the too-low ceiling. The fix was the constructive-code floor.

After these three fixes, a second billed run showed G3 up from 54\% to 96\% (a pass) but G2 unchanged at 43\%: the same earnest turns kept revising, and the run did not record \emph{why}. We added reason capture (\S\ref{sec:diag}) and ran again.

\textbf{Rung 3, fabricated citations.} With reasons in hand, the residual G2 failures split. Two-thirds were a subtle grounding failure: the contract told the actor to ``ground every course claim in the retrieved material,'' which pressured it to cite, and when nothing relevant had been retrieved it \emph{invented} a plausible citation. The fix: cite a specific source only if it appears in this turn's retrieval, otherwise help in plain words. The remaining third was code over the ceiling, because ``here is my code, what is wrong?'' was labeled a question rather than an attempt, so the constructive-code floor never applied; the fix extended the floor to any code-adjacent turn.

\textbf{Rung 4, naming the bug, and over-grounding general facts.} These fixes drove the earnest revise rate to 0\%, but two further runs exposed the subtlest over-help yet. On a debugging turn the actor would stay within the no-code ceiling but \emph{state the exact bug in prose} (``you wrote less-than-or-equal where you need less-than''), which the auditor correctly flagged as giving the answer; the fix sharpened the low-rung contract (point to the area and ask one leading question, never naming the bug, the wrong token, or the fix). Then one earnest turn still revised because the judge flagged a \emph{general} language fact (what a standard container's size method returns) as an unsupported course claim; the fix narrowed the rule to treat general programming facts as the tutor's own knowledge.

\textbf{End state.} The final run passed both pedagogy gates: a 0\% earnest revise rate (G2) and 100\% hint-ceiling compliance, with zero rung breaches under adversarial pressure (G3), at a driven cost well under a dollar. With the always-on deterministic gates (G1 zero reveals, G4 zero exam compromises), all four acceptance criteria held. The progression is the point: the failures walked steadily \emph{down} an over-help ladder, each rung subtler than the last, every one caught and named by the captured judge reason and fixed by cause until none remained. No single prompt edit produced a withholding tutor; the descent down the ladder did.

\section{Discussion and reflection}
\label{sec:disc}

\textbf{Refusal-under-knowledge is distinct and under-studied.} Nothing in the calibration loop is specific to data structures, or even to education. Any LLM agent whose core requirement is to \emph{refuse a capability it possesses} faces the same failure modes: the model edges toward the forbidden behavior, and naive metrics conflate distinct failures. Classical LLM safety studies refusal of content the model should not produce; this is the harder converse, refusing content it should and can produce, to a user who can see that it can. The over-help ladder is the tutoring instance of a general phenomenon: a model's violations of a withholding contract are ordered gross-to-subtle, and a reason-capturing loop walks down that order.

\textbf{Put the irreversible decision in code, not in a prompt.} The single most stabilizing choice was P2: computing the help ceiling in a non-LLM policy core fed only trusted state made the most important property both injection-proof and unit-testable, so prompt edits could never silently weaken it. Equally valuable was honest tooling: naming the detector's blind spots and forcing the auditor to state its reason caught errors a green pass rate would have hidden, including an auditor itself wrong in a structured way (Rung 0). For a safety-relevant system, instrumentation that explains itself proved more valuable than any single guardrail.

\textbf{Use of generative AI in this work.} Per ACM authorship policy, we disclose the AI used in conducting this research. The deployed system uses LLMs as runtime components (the actor, pre-classifier, strategist, and judge) and a stronger LLM as the offline auditor, described in \S\ref{sec:arch} and~\S\ref{sec:method}; the models are used as-is through their providers' APIs with no fine-tuning, and all corpora used for persona grounding and retrieval are public-domain or instructor-authored. Generative AI tools also assisted in writing system code and in drafting and editing this manuscript; the author verified all resulting content and is responsible for any errors.

\section{Limitations}
\label{sec:limits}

This report validates an enforced-pedagogy system against safety and compliance gates; it does not claim a measured durable-learning effect, which requires the controlled study in~\S\ref{sec:future}. The pedagogy gates pass on roughly two dozen driven turns run a handful of times across the scripted personas, which does not substitute for real student traffic (monitored by the nightly auditor in a pilot). Both the live judge and the auditor are LLMs and inherit the known reliability limits of LLM-as-judge, including for programming feedback~\cite{zheng2023judge,koutcheme2025feedback}; we mitigate with the rubric, deterministic detector, and escalation described above, plus planned calibration against human raters. The detector misses split-fence, prose, and cross-language reveals by design, with the judge and execution path as compensating layers; since a determined adversary can always retype the problem elsewhere, the goal is not leak-proofness but making the guarded path the path of least resistance. With real students, conversation logs and code submissions are sensitive data, and any research analysis is gated behind human-subjects review. Finally, the persona carries risks (an idealized voice may be over-trusted), so it is subordinated to clarity and the system discloses its AI status.

\section{Future work}
\label{sec:future}

The immediate next step is a controlled, pre-registered study in one course offering, with a delayed, tool-removed assessment on goal-aligned items two weeks after a unit, contrasting no tutor, the supervised tutor, and an instructor-enabled explain-mode arm to isolate the effect of withholding itself~\cite{bastani2024genai}. This study involves human participants and will proceed only under institutional human-subjects approval. Secondary measures include depth to convergence, voluntary-usage persistence, and an equity slice by incoming ability, since the intervention should narrow rather than widen ability gaps~\cite{heffernan2014assistments}. We will also adopt forced-choice pairwise preference judging by TAs and the instructor, shown affordable at scale~\cite{liu2024cs50}, to A/B the persona and pedagogy levers, and measure how much guardrail friction students tolerate before abandoning the guarded tool, extending early evidence on optional guardrails~\cite{kapoor2025guardrails} to the enforced-contract setting.

\section{Conclusion}

A useful LLM tutor must do what a capable model finds hardest: withhold an answer it knows, under pressure, while still helping. We have shown this is achievable, but not by prompt-writing. It requires an architecture that makes withholding an enforced, per-turn contract decided outside the generating model and backed by a deterministic check, plus a method that calibrates the behavior against evidence and makes each rejection state its reason. Applied to a deployed tutor, the failures descended an interpretable over-help ladder to full compliance on all four gates. That measure-diagnose-fix loop is the contribution, and should transfer to any agent that must refuse a capability it has.

\begin{acks}
Generative AI tools (large language models) were used substantially in this
work: as runtime components of the deployed tutoring system, and as assistants
in writing the system's code and in drafting and editing this manuscript. The
author verified all content, results, and references and is responsible for any
errors. See ``Use of generative AI in this work'' (\S\ref{sec:disc}) for the
full disclosure.
\end{acks}

\balance
\bibliographystyle{ACM-Reference-Format}
\bibliography{refs}

\end{document}